\documentclass[letterpaper,USenglish]{lipics}
\usepackage{amsfonts}
\usepackage{amsmath}

\usepackage{microtype}

\title{Strong converse for the quantum capacity of the erasure channel for
almost all codes} 
\titlerunning{Strong converse for the quantum capacity of the erasure channel for
almost all codes} 

\author[1]{Mark M. Wilde}
\author[2]{Andreas Winter}
\affil[1]{Hearne Institute for Theoretical Physics\\Department of
Physics and Astronomy\\Center for Computation and Technology\\Louisiana State
University\\Baton Rouge, Louisiana 70808, USA\\
  \texttt{mwilde@lsu.edu}
  }
\affil[2]{ICREA \& F\'isica Te\'orica\\Informaci\'o i Fenomens
Qu\'antics\\ Universitat Aut\`{o}noma de Barcelona\\ ES-08193 Bellaterra
(Barcelona), Spain\\
  \texttt{andreas.winter@uab.cat}
  }
\authorrunning{M. M. Wilde and A. Winter} 
\Copyright{M. M. Wilde and A. Winter}
\subjclass{H.1.1 Systems and Information Theory, E.4 Coding and Information Theory, Error control codes}
\keywords{strong converse, quantum erasure channel, quantum capacity}
\serieslogo{}
\volumeinfo
  {}
  {2}
  {$9^{\text{th}}$ Conference on the Theory of Quantum Computation, Communication, and Cryptography}
  {1}
  {1}
  {1}
\EventShortName{}
\DOI{10.4230/LIPIcs.xxx.yyy.p}

\let\originalleft\left
\let\originalright\right
\def\left#1{\mathopen{}\originalleft#1}
\def\right#1{\originalright#1\mathclose{}}

\newcommand{\ket}[1]{|#1\rangle}
\newcommand{\cN}{\mathcal{N}}

\begin{document}

\maketitle

\begin{abstract}
A strong converse theorem for channel capacity establishes that the error probability in any communication
scheme for a given channel necessarily tends to one if the rate of communication exceeds the channel's capacity. Establishing
such a theorem for the quantum capacity of degradable channels has been an elusive task, with the
strongest progress so far being a so-called ``pretty strong converse.'' In this work, Morgan and Winter
proved that the quantum error of any quantum communication scheme for a given degradable channel converges to a value larger than $1/\sqrt{2}$ in the limit of many channel uses
if the quantum rate of communication exceeds the channel's quantum capacity. The present paper establishes a theorem that is a counterpart to this ``pretty strong converse.'' We prove
that the large fraction of codes having a rate exceeding the erasure channel's quantum capacity have a quantum error tending to one in the limit of many channel uses. Thus, our work adds to the body of evidence that a fully strong converse theorem should hold for the quantum capacity of the erasure channel. As a side result, we prove that the classical capacity of the quantum erasure channel obeys the strong converse property.
\end{abstract}

\section{Introduction}

In his seminal paper on quantum error correction, Shor set out the task of
determining the quantum capacity of a quantum channel \cite{S95}, defined as
the maximum rate at which it is possible to transmit qubits reliably over a
noisy quantum communication channel. Subsequent to this, the coherent
information was identified as being a relevant quantity for quantum capacity 
\cite{SN96}, a regularized upper bound on quantum capacity was established
in terms of the coherent information \cite{BKN98,BNS98}, and the coherent
information lower bound on the quantum capacity was established by a
sequence of works which are often said to bear \textquotedblleft increasing
standards of rigor\textquotedblright\ \cite{L97,capacity2002shor,D05}.%
\footnote{%
However, see the later works in \cite{qcap2008second} and \cite%
{qcap2008fourth}, which respectively set \cite{L97} and \cite%
{capacity2002shor} on a firm foundation.} All of these works did not
identify a tractable characterization of the quantum capacity in general,
but Devetak and Shor subsequently proved that the coherent information is
equal to the quantum capacity for a class of channels bearing the property
of degradability \cite{cmp2005dev}. Degradable channels are such that the
receiver of the output of the channel can simulate the channel to the
environment by applying a degrading map.

A particularly simple example of a degradable channel is the quantum erasure
channel~$\mathcal{N}_{p}$\ \cite{GBP97}, which has the following action on
an input density operator $\rho$:%
\begin{equation}
\mathcal{N}_{p}\left( \rho\right) \equiv\left( 1-p\right) \rho+p\left\vert
e\right\rangle \left\langle e\right\vert ,   \label{eq:erasure-channel}
\end{equation}
where $p\in\left[ 0,1\right] $ is the erasure probability and $\left\vert
e\right\rangle $ is a state orthogonal to the input space (i.e., $%
\left\langle e\right\vert \rho\left\vert e\right\rangle =0$ for all input $%
\rho$). One can readily show that the map to the environment is equivalent
(up to isometry) to an erasure channel with the complementary probability:%
\begin{equation}
\mathcal{N}_{p}\left( \rho\right) \equiv p\rho+\left( 1-p\right) \left\vert
e\right\rangle \left\langle e\right\vert .
\end{equation}
The interpretation here is that if the receiver recovers the channel input,
then the environment does not and instead receives the erasure flag, and
vice versa.

The quantum capacity of the erasure channel was identified early on by
employing a now well known \textquotedblleft no-cloning\textquotedblright\
argument \cite{PhysRevLett.78.3217}. That is, when $p=1/2$, the channels
from input to the receiver and from input to the environment are the same,
so that the quantum capacity of the original channel must vanish. If this
were not the case, then it would be possible to send quantum data reliably
to both the receiver and the environment of the channel, in violation of the
no-cloning theorem. It is then possible to prove that the quantum capacity
of the erasure channel in general is equal to $\left( 1-2p\right) \log d$
for $p\geq1/2$ and zero otherwise (in agreement with the aforementioned
reasoning), where $d$ is the dimension of the input space for the channel.

All of the above works established an understanding of quantum capacity in
the following sense:
\begin{enumerate}
\item (Achievability)\ If the rate of quantum communication is below the
quantum capacity, then there exists a scheme for quantum communication such
that the fidelity approaches one in the limit of many channel uses.

\item (Weak Converse) If the rate of quantum communication is above the
quantum capacity, then there cannot exist an error-free quantum
communication scheme.
\end{enumerate}
However, the theorem stated as such still leaves more to be desired. For
example, it has been known for a long time that the classical capacity of a
classical channel obeys the strong converse property \cite%
{Wolfowitz1964,Arimoto73}:\ if the rate of communication exceeds capacity,
then the error probability necessarily converges to one in the limit of many
channel uses. Furthermore, many works have now established that the strong
converse property holds for the classical capacity of several quantum
channels \cite{W99,Ogawa,KW09,WWY13,WW13,BPWW14} and for the
entanglement-assisted classical capacity of all quantum channels \cite%
{BDHSW12,BCR09,GW13}.

Thus, we are left with the strong converse question for the quantum
capacity, with the goal being to sharpen our understanding of quantum
capacity. In general, the quantum capacity of arbitrary channels can exhibit
rather exotic behavior \cite{science2008smith}, so it seems reasonable to
restrict attention for now to the class of degradable channels since they
are more well behaved. In this spirit, a recent work has proved that the
quantum capacity of all degradable channels exhibits a property dubbed the
\textquotedblleft pretty strong converse\textquotedblright\ \cite{MW13}.
These authors have proven that the quantum error\footnote{%
As quantified by the so-called \textquotedblleft purified
distance\textquotedblright\ (see Chapter~3 of \cite{T12}, for example).} of
any quantum communication scheme for a degradable channel experiences a
sudden jump from zero to at least $1/\sqrt{2}$ when the communicate rate crosses the
quantum capacity threshold (this statement is in the limit of many channel
uses). At the very least, we now know that the quantum capacity experiences
this jump, but the work of \cite{MW13} left open the question of whether the
jump in quantum error is actually from zero to one in the limit of many
channel uses.

In this paper, we prove a statement that is similar in spirit to the pretty
strong converse:\ for almost all codes having a rate exceeding the quantum
capacity of the erasure channel, the error necessarily converges to one in
the limit of many channel uses. We should clarify that we do not prove a
strong converse for all codes, but instead show that the strong converse
property holds for almost all codes. We will be more precise in what follows
with clarifying what we mean by \textquotedblleft almost all
codes,\textquotedblright\ but suffice it for now to say if anyone devises a
communication scheme for quantum communication over the erasure channel
whose rate exceeds capacity, then the chances are very good that, regardless
of the scheme, it will fail with probability converging to one in the limit
of many channel uses.

In the absence of a proof that the strong converse holds, both the present
paper and \cite{MW13} are offering an increasing body of evidence that it
should indeed hold for the class of quantum erasure channels. That is, both
results allow us to conclude the following statement:\ all codes whose rate
exceeds the quantum capacity of the erasure channel have a quantum error
converging to $1/\sqrt{2}$ in the limit of many channel uses, and a large
fraction of them in fact have quantum error converging to one.

This paper is organized as follows. The next section reviews the definition
of an entanglement generation code. Section~\ref{sec:gen-div} then reviews
the generalized divergence framework of Sharma and Warsi \cite{SW12}\ for
establishing bounds relating rate, error, and the channel of interest in any
quantum communication protocol. Section~\ref{sec:main-result} provides
a proof for our main result:\ that the strong converse property holds for
almost all codes used for quantum communication over the quantum erasure
channel. We state some open directions in the conclusion. The appendix
includes, as a side result, a proof that the strong converse holds for the
classical capacity of the quantum erasure channel.

\section{Entanglement generation codes}

In this paper, we focus on entanglement generation codes, for which the goal
is for the sender Alice to use the channel $n$ times in order to share a
state with the receiver Bob, such that this state is indistinguishable from
a maximally entangled state. We focus on this task because the entanglement
generation capacity of a quantum channel serves as an upper bound on its
quantum capacity (this in turn is because a protocol for noiseless quantum
communication can always be used to generate entanglement between sender and
receiver). Thus, if one establishes an upper bound on the entanglement
generation capacity, then this bound serves as an upper bound on the quantum
capacity. However, we should emphasize again that our final statement is a
bound that holds for almost all entanglement generation codes, so that we
cannot conclude a full strong converse.

More formally, we now define an $\left( n,R,\varepsilon,\phi,D\right) $
entanglement generation code for a channel~$\mathcal{N}$. Such a protocol
begins with Alice preparing a state on $n+1$ systems, she sends $n$ shares
of the state through $n$ instances of the channel, and then Bob decodes.
That is, such a code begins with Alice preparing a state $\left\vert
\phi\right\rangle _{AA_{1}\cdots A_{n}}$. The reduced state on system $A$
has its rank equal to $M$, where $M=2^{nR}$. Alice then transmits systems $%
A_{1}\cdots A_{n}$ through $n$ uses of the channel, leading to the state%
\begin{equation}
\rho_{AB^{n}}\equiv\mathcal{N}_{A^{n}\rightarrow B^{n}}\left( \phi
_{AA_{1}\cdots A_{n}}\right) ,   \label{eq:state-output-from-channel}
\end{equation}
where $\mathcal{N}_{A^{n}\rightarrow B^{n}}\equiv\mathcal{N} ^{\otimes n}$
and $A^{n}$ is shorthand for $A_{1}\cdots A_{n}$. Finally, Bob performs a
decoding $D_{B^{n}\rightarrow\hat{B}}$, leading to the state%
\begin{equation}
\omega_{A\hat{B}}\equiv D_{B^{n}\rightarrow\hat{B}}\left( \mathcal{N}%
_{A^{n}\rightarrow B^{n}}\left( \phi_{AA_{1}\cdots A_{n}}\right) \right) .
\end{equation}
The fidelity of the code is given by%
\begin{equation}
F\equiv\left\langle \Phi\right\vert _{A\hat{B}}\ \omega_{A\hat{B}}\
\left\vert \Phi\right\rangle _{A\hat{B}},   \label{eq:fidelity}
\end{equation}
where $\left\vert \Phi\right\rangle _{A\hat{B}}$ is the maximally entangled
state%
\begin{equation}
\left\vert \Phi\right\rangle _{A\hat{B}}\equiv\frac{1}{\sqrt{M}}\sum
_{i=0}^{M-1}\left\vert i\right\rangle _{A}\left\vert i\right\rangle _{\hat{B}%
},
\end{equation}
so that the rate of entanglement generation is equal to $\frac{1}{n}\log_{2}M
$. An $\left( n,R,\varepsilon,\phi,D\right) $ code uses the state $\phi$,
the decoder $D$, the channel $n$ times at rate $R$, and is such that the
fidelity $F\geq1-\varepsilon$. Note that without loss of generality, we can
restrict our consideration to pure-state entanglement generation codes. For
if the initial state is a mixed state $\rho_{AA_{1}\cdots A_{n}}$ and the
following condition holds%
\begin{equation}
\left\langle \Phi\right\vert _{A\hat{B}} D_{B^{n}\rightarrow\hat{B}}\left( 
\mathcal{N}_{A^{n}\rightarrow B^{n}}\left( \rho_{AA_{1}\cdots A_{n}}\right)
\right) \left\vert \Phi\right\rangle _{A\hat{B}}\geq1-\varepsilon,
\end{equation}
then there always exists at least one pure state in the spectral
decomposition of $\rho_{AA_{1}\cdots A_{n}}$ which meets the same fidelity
constraint given above.

\section{Generalized divergence framework for quantum communication}

\label{sec:gen-div}

We now recall the Sharma-Warsi framework for bounding fidelities in quantum
communication \cite{SW12}. We say that $\mathcal{D}\left( X||Y\right) $ is a 
\textit{generalized divergence} if it satisfies the following monotonicity
inequality for all quantum channels $\mathcal{M}$ and positive operators $X$
and $Y$:%
\begin{equation}
\mathcal{D}\left( X||Y\right) \geq\mathcal{D}\left( \mathcal{M}\left(
X\right) ||\mathcal{M}\left( Y\right) \right) .
\end{equation}
Let $I_{\mathcal{D}}\left( A\rangle B\right) _{\rho}$ denote the generalized
coherent information of a bipartite state $\rho_{AB}$:%
\begin{equation}
I_{\mathcal{D}}\left( A\rangle B\right) _{\rho}\equiv\min_{\sigma_{B}}%
\mathcal{D}\left( \rho_{AB}||I_{A}\otimes\sigma_{B}\right) .
\end{equation}
Let $I_{\mathcal{D}}\left( \mathcal{N}\right) $ denote the generalized
coherent information of a quantum channel $\mathcal{N}$:%
\begin{align}
I_{\mathcal{D}}\left( \mathcal{N}\right) & \equiv\max_{\phi_{AA^{\prime}}}I_{%
\mathcal{D}}\left( A\rangle B\right) _{\mathcal{N}_{A^{\prime }\rightarrow
B}\left( \phi_{AA^{\prime}}\right) }  \label{eq:gen-coh-info} \\
& =\max_{\phi_{AA^{\prime}}}\min_{\sigma_{B}}\mathcal{D}\left( \mathcal{N}%
_{A^{\prime}\rightarrow B}\left( \phi_{AA^{\prime}}\right)
||I_{A}\otimes\sigma_{B}\right) .
\end{align}
If the generalized divergence is equal to the von Neumann relative entropy,
then the above expressions are equal to the usual coherent information of a
quantum state and coherent information of a quantum channel, respectively.

We now establish a bound relating the rate and error of any entanglement
generation code for a quantum channel $\mathcal{N}$\ to the generalized
coherent information of the tensor-power channel $\mathcal{N}^{\otimes n}$.
For our purposes here, we begin by considering the generalized divergence
between the state $\rho_{AB^{n}}$ defined in (\ref%
{eq:state-output-from-channel}) that is output from $n$ uses of the channel
and any other operator of the form $I_{A}\otimes\sigma_{B^{n}}$, where $%
\sigma_{B^{n}}$ is a density operator on the systems $B^{n}$:%
\begin{equation}
\mathcal{D}\left( \rho_{AB^{n}}||I_{A}\otimes\sigma_{B^{n}}\right) .
\end{equation}
By monotonicity under the application of the decoder $D_{B^{n}\rightarrow 
\hat{B}}$\ to the system $B^{n}$, the following inequality holds%
\begin{equation}
\mathcal{D}\left( \rho_{AB^{n}}||I_{A}\otimes\sigma_{B^{n}}\right) \geq%
\mathcal{D}\left( \omega_{A\hat{B}}||I_{A}\otimes D_{B^{n}\rightarrow \hat{B}%
}\left( \sigma_{B^{n}}\right) \right) .
\end{equation}
Next, consider the following test (a completely positive trace-preserving\
map), which outputs a flag indicating whether a state is maximally entangled
or not:%
\begin{equation}
T_{A\hat{B}\rightarrow Z}\left( \cdot\right) \equiv\text{Tr}\left\{ \Phi_{A%
\hat{B}}\left( \cdot\right) \right\} \left\vert 1\right\rangle \left\langle
1\right\vert +\text{Tr}\left\{ \left( I_{A\hat{B}}-\Phi _{A\hat{B}}\right)
\left( \cdot\right) \right\} \left\vert 0\right\rangle \left\langle
0\right\vert .
\end{equation}
Intuitively, this test is simply asking, \textquotedblleft Is the
entanglement decoded or not?\textquotedblright\ Applying monotonicity of the
generalized divergence under this test, we find that the following
inequality holds%
\begin{equation}
\mathcal{D}\left( \omega_{A\hat{B}}||I_{A}\otimes D_{B^{n}\rightarrow\hat{B}%
}\left( \sigma_{B^{n}}\right) \right) \geq\mathcal{D}\left( T_{A\hat {B}%
\rightarrow Z}\left( \omega_{A\hat{B}}\right) ||T_{A\hat{B}\rightarrow
Z}\left( I_{A}\otimes D_{B^{n}\rightarrow\hat{B}}\left(
\sigma_{B^{n}}\right) \right) \right) .
\end{equation}
By defining%
\begin{align}
\rho_{F} & \equiv F\left\vert 1\right\rangle \left\langle 1\right\vert
+\left( 1-F\right) \left\vert 0\right\rangle \left\langle 0\right\vert , \\
P_{\frac{1}{M}} & \equiv\frac{1}{M}\left\vert 1\right\rangle \left\langle
1\right\vert +\left( M-\frac{1}{M}\right) \left\vert 0\right\rangle
\left\langle 0\right\vert ,
\end{align}
we see that%
\begin{equation}
\mathcal{D}\left( T_{A\hat{B}\rightarrow Z}\left( \omega_{A\hat{B}}\right)
||T_{A\hat{B}\rightarrow Z}\left( I_{A}\otimes D_{B^{n}\rightarrow\hat{B}%
}\left( \sigma_{B^{n}}\right) \right) \right) =\mathcal{D}\left( \rho
_{F}||P_{\frac{1}{M}}\right) ,
\end{equation}
which follows from (\ref{eq:fidelity}) and the fact that%
\begin{equation}
\text{Tr}\left\{ \Phi_{A\hat{B}}\left( I_{A}\otimes D_{B^{n}\rightarrow \hat{%
B}}\left( \sigma_{B^{n}}\right) \right) \right\} =\frac{1}{M}.
\end{equation}
Thus, putting everything together, we obtain the following inequality%
\begin{equation}
\mathcal{D}\left( \rho_{AB^{n}}||I_{A}\otimes\sigma_{B^{n}}\right) \geq%
\mathcal{D}\left( \rho_{F}||P_{\frac{1}{M}}\right) .
\end{equation}
This inequality holds for any choice of $\sigma_{B^{n}}$, so we can obtain
the tightest upper bound on $\mathcal{D(}\rho_{F}||P_{\frac{1}{M}})$\ for a
particular entanglement generation code with initial state $%
\phi_{AA_{1}\cdots A_{n}}$ by taking a minimization over all such $%
\sigma_{B^{n}}$:%
\begin{equation}
\min_{\sigma_{B^{n}}}\mathcal{D}\left( \rho_{AB^{n}}||I_{A}\otimes
\sigma_{B^{n}}\right) \geq\mathcal{D}\left( \rho_{F}||P_{\frac{1}{M}}\right)
.   \label{eq:gen-div-bound}
\end{equation}
We can then remove the dependence of the bound on any particular
entanglement generation code by taking a maximization over all initial
states $\phi _{AA_{1}\cdots A_{n}}$:%
\begin{equation}
\max_{\phi_{AA_{1}\cdots A_{n}}}\min_{\sigma_{B^{n}}}\mathcal{D}\left(
\rho_{AB^{n}}||I_{A}\otimes\sigma_{B^{n}}\right) \geq\mathcal{D}\left(
\rho_{F}||P_{\frac{1}{M}}\right) .
\end{equation}
By employing the definition in (\ref{eq:gen-coh-info}), we find that the
bound is equivalent to%
\begin{equation}
I_{\mathcal{D}}\left( \mathcal{N}^{\otimes n}\right) \geq\mathcal{D}\left(
\rho_{F}||P_{\frac{1}{M}}\right) .
\end{equation}

\subsection{Specializing to R\'{e}nyi relative entropies}

The above development applies for any divergence satisfying monotonicity,
and the R\'{e}nyi relative entropy is a particular example of a generalized
divergence, defined as%
\begin{equation}
D_{\alpha}\left( \rho||\sigma\right) \equiv\frac{1}{\alpha-1}\log _{2}\text{%
Tr}\left\{ \rho^{\alpha}\sigma^{1-\alpha}\right\} .
\end{equation}
Monotonicity of $D_{\alpha}\left( \rho||\sigma\right) $ under quantum
channels holds for all $\alpha\in\left[ 0,2\right] $ (see Appendix~B of \cite%
{T12}, for example). In the present paper, we are focused on $\alpha \in(1,2]
$, especially when $\alpha$ is in a neighborhood near one in this interval. This is because
the R\'enyi relative entropy converges to the von Neumann relative entropy as
$\alpha \to 1$.

Now we can evaluate the bound in (\ref{eq:gen-div-bound})\ for the case when
the divergence is chosen to be the R\'{e}nyi relative entropy:%
\begin{align}
\min_{\sigma_{B^{n}}}\mathcal{D}\left( \rho_{AB^{n}}||I_{A}\otimes
\sigma_{B^{n}}\right) & \geq D_{\alpha}\left( \rho_{F}||P_{\frac{1}{M}%
}\right)  \label{eq:bound-1} \\
& =\frac{1}{\alpha-1}\log_{2}\left[ F^{\alpha}\left( \frac{1}{M}\right)
^{1-\alpha}+\left( 1-F\right) ^{\alpha}\left( M-\frac{1}{M}\right)
^{1-\alpha}\right] \\
& \geq\frac{1}{\alpha-1}\log_{2}\left[ F^{\alpha}\left( \frac{1}{M}\right)
^{1-\alpha}\right] \\
& =\frac{\alpha}{\alpha-1}\log_{2}\left[ F\right] +\log_{2}M \\
& =\frac{\alpha}{\alpha-1}\log_{2}\left[ F\right] +nR   \label{eq:bound-last}
\end{align}
If we optimize over all entanglement generation codes, then we have the bound%
\begin{equation}
\max_{\phi_{AA_{1}\cdots A_{n}}}\min_{\sigma_{B^{n}}}D_{\alpha}\left(
\rho_{AB^{n}}||I_{A}\otimes\sigma_{B^{n}}\right) \geq\frac{\alpha}{\alpha -1}%
\log_{2}\left[ F\right] +nR.
\end{equation}
This is equivalent to%
\begin{equation}
I_{\alpha}\left( \mathcal{N}^{\otimes n}\right) \geq\frac{\alpha}{\alpha -1}%
\log_{2}\left[ F\right] +nR,
\end{equation}
where we define the R\'{e}nyi coherent information $I_{\alpha}$ of a quantum
channel according to the recipe in (\ref{eq:gen-coh-info}). Rewriting this,
the bound is equivalent to%
\begin{equation}
F\leq2^{-n\left( \frac{\alpha-1}{\alpha}\right) \left( R-\frac{1}{n}%
I_{\alpha}\left( \mathcal{N}^{\otimes n}\right) \right) }.
\end{equation}

\begin{remark}
It is worth noting at this point that if it is possible to prove that
$\frac{1}{n}I_{\alpha}\left(  \mathcal{N}^{\otimes n}\right)  $ is an additive
function of the channel $\mathcal{N}$, in the sense that%
\begin{equation}
\frac{1}{n}I_{\alpha}\left(  \mathcal{N}^{\otimes n}\right)  =I_{\alpha
}\left(  \mathcal{N}\right)
\end{equation}
for any finite $n$, then this would be sufficient to prove that the strong
converse holds according to the argument of \cite{Ogawa}\ (which has since
been repeated in different contexts in both \cite{SW12}\ and \cite{GW13}). (In
fact, any subadditivity relation of the following form would
suffice:\ $I_{\alpha}\left(  \mathcal{N}^{\otimes n}\right)  \leq nI_{\alpha
}\left(  \mathcal{N}\right)  +o\left(  n\right)  $.) One could also consider
using the recently developed sandwiched R\'{e}nyi relative entropy
\cite{MDSFT13,WWY13} in this context. So far, it is not clear to us whether
either of the coherent information quantities derived from the traditional or
sandwiched R\'{e}nyi relative entropies are additive in the above sense for
any degradable channel.
\end{remark}

%QUESTION:\ should we have a discussion about our findings regarding
%non-additivity for the erasure channel?

\subsection{Application to the quantum erasure channel}

\label{sec:erasure-channel-app}We now specialize the above bounds to the
case of the quantum erasure channel. Beginning from (\ref{eq:bound-1})-(\ref%
{eq:bound-last}), we see that we can choose any state $\sigma_{B^{n}}$ for
establishing a bound relating rate and fidelity to an information quantity.
So we choose $\sigma_{B^{n}}=\left[ \mathcal{N}_{p}\left( \pi\right) \right]
^{\otimes n}=((1-p)\pi+p|e\rangle\langle e|)^{\otimes n}$, where $\pi=I/d$
is the maximally mixed qudit state on the input and $\mathcal{N}_{p}$ is the
erasure channel defined in (\ref{eq:erasure-channel}). This then leads to
the following bound for any $\left( n,R,\varepsilon ,\phi,D\right) $
entanglement generation code:%
\begin{align}
\frac{\alpha}{\alpha-1}\log_{2}\left[ F\left( \phi\right) \right] +nR &
\leq\min_{\sigma_{B^{n}}}{D}_{\alpha}\left( \mathcal{N}_{A\rightarrow
B^{n}}\left( \phi_{AA_{1}\cdots A_{n}}\right)
||I_{A}\otimes\sigma_{B^{n}}\right)  \label{eq:fidelity-bound-erasure} \\
& \leq{D}_{\alpha}\left( \mathcal{N}_{A^{n}\rightarrow B^{n}}\left(
\phi_{AA_{1}\cdots A_{n}}\right) ||I_{A}\otimes\left[ \mathcal{N}_{p}\left(
\pi\right) \right] ^{\otimes n}\right) ,
\end{align}
where $\mathcal{N}_{A^{n}\rightarrow B^{n}}=\mathcal{N}_{p}^{\otimes n}$ and 
$F\left( \phi\right) $ denotes the fidelity of an entanglement generation
code with initial state $\phi$.\footnote{%
We could denote this fidelity as $F\left( \phi,D\right) $ because the
fidelity of any code depends on the initial state $\phi$ and the decoder $D$%
, but the bound we find here is independent of the decoder $D$, so we
suppress it from the notation.} Observe now that the output of $n$ uses of
the quantum erasure channel is rather special, in the sense that it can be
written as a convex combination of $2^{n}$ density operators which are
supported on orthogonal subspaces. We can index these by a binary string $i$
(where ones in this string represent the systems that get erased and zeros
represent systems that do not get erased), and we denote the density
operators for $\mathcal{N}_{A^{n}\rightarrow B^{n}}\left( \phi_{AA_{1}\cdots
A_{n}}\right) $ by $\omega_{AB^{n}}^{i}$ and those for $\left[ \mathcal{N}%
\left( \pi\right) \right] ^{\otimes n}$ by $\tau_{B^{n}}^{i}$. Furthermore,
let $\left\{ i\right\} \ $be the set of indices for the systems that get
erased, so that we denote the systems that get erased by $A^{\left\{
i\right\} }$ and those that do not by $A^{\left\{ i\right\} ^{c}}$. We then
find that%
\begin{align}
& {D}_{\alpha}\left( \mathcal{N}_{A^{n}\rightarrow B^{n}}\left( \phi
_{AA_{1}\cdots A_{n}}\right) ||I_{A}\otimes\left[ \mathcal{N}_{p}\left(
\pi\right) \right] ^{\otimes n}\right)  \nonumber \\
& =\frac{1}{\alpha-1}\log\sum_{i\in\left\{ 0,1\right\} ^{n}}\left(
1-p\right) ^{n-\left\vert i\right\vert }p^{\left\vert i\right\vert }\text{Tr}%
\left\{ \left[ \omega_{AB^{n}}^{i}\right] ^{\alpha}\left( I_{A}\otimes\left(
\tau_{B^{n}}^{i}\right) ^{1-\alpha}\right) \right\} \\
& =\frac{1}{\alpha-1}\log\sum_{i\in\left\{ 0,1\right\} ^{n}}\left[ \left(
1-p\right) d^{\alpha-1}\right] ^{n-\left\vert i\right\vert }p^{\left\vert
i\right\vert }\text{Tr}\left\{ \left[ \phi_{AA^{\left\{ i\right\} ^{c}}}%
\right] ^{\alpha}\right\} \\
& =\frac{1}{\alpha-1}\log\sum_{i\in\left\{ 0,1\right\} ^{n}}\left[ \left(
1-p\right) d^{\alpha-1}\right] ^{n-\left\vert i\right\vert }p^{\left\vert
i\right\vert }\text{Tr}\left\{ \left[ \phi_{A^{\left\{ i\right\} }}\right]
^{\alpha}\right\} ,   \label{eq:rewrite-renyi-erasure}
\end{align}
where the last equality follows because the spectrum of $\phi_{AA^{\left\{
i\right\} ^{c}}}$ is equal to the spectrum of $\phi_{A^{\left\{ i\right\} }}$
for a pure state. Rewriting (\ref{eq:fidelity-bound-erasure})-(\ref%
{eq:rewrite-renyi-erasure}), we obtain the following bound on the fidelity~$%
F\left( \phi\right) $:%
\begin{equation}
F\left( \phi\right) \leq\left[ 2^{-n\left( \frac{\alpha-1}{\alpha}\right) R}%
\right] \Bigg[ \sum_{i\in\left\{ 0,1\right\} ^{n}}\left[ \left( 1-p\right)
d^{\alpha-1}\right] ^{n-\left\vert i\right\vert }p^{\left\vert i\right\vert }%
\text{Tr}\left\{ \left[ \phi_{A^{\left\{ i\right\} }}\right]
^{\alpha}\right\} \Bigg] ^{\frac{1}{\alpha}}. 
\label{eq:fidelity-bound-for-all}
\end{equation}

\begin{remark}
By inspecting the above, we see that obtaining a general bound on the fidelity
of an entanglement generation code for the quantum erasure channel is related
to the quantum marginal problem \cite{K04}, since the various terms
$\operatorname{Tr}\left\{  \left[  \phi_{A^{\left\{  i\right\}  }}\right]
^{\alpha}\right\}  $ in the sum are the $\alpha$-purities of all of the
$2^{n}$\ marginals of the quantum state $\phi_{AA_{1}\cdots A_{n}}$.
\end{remark}

\section{Strong converse for almost all codes}

\label{sec:main-result}

In the previous section, we established the bound (\ref%
{eq:fidelity-bound-for-all}) on the fidelity $F\left( \phi\right) $ of any $%
\left( n,R,\varepsilon,\phi,D\right) $\ entanglement generation code. In
this section, we prove our main result, i.e., that the large fraction of
capacity-exceeding entanglement generation codes satisfy the strong converse
property. Before proving this result, we need to establish a measure on the
set of all entanglement generation codes, in order to talk about the
fraction of codes that satisfy the strong converse property. The most
natural measure in this context is the unitarily invariant measure (Haar
measure) on pure states, so that each possible initial state for an
entanglement generation code is \textquotedblleft receiving equal
weight.\textquotedblright

Now, suppose that we select the pure state $\phi _{AA^{ n}}$ at random
according to the Haar measure with $\left\vert A\right\vert =2^{nR}$ and $%
\left\vert A_{i}\right\vert =d$ for all $i\in \left\{ 1,\ldots ,n\right\} $.
What makes the subsequent reasoning pertinent is the well-known fact
that for $R < Q(\mathcal{N}_p) = (1-2p)\log d$, this choice results in a good
code asymptotically with overwhelming probability. 
(Cf.~for instance \cite{qcap2008fourth}.)

We begin by analyzing the expectation of the fidelity $F\left( \phi \right)$:
\begin{align}
\mathbb{E}_{\phi }\left\{ F\left( \phi \right) \right\} & \leq \mathbb{E}%
_{\phi }\left\{ 2^{-n\left( \frac{\alpha -1}{\alpha }\right) R}\bigg[
\sum_{i\in \left\{ 0,1\right\} ^{n}}\left[ \left( 1-p\right) d^{\alpha -1}%
\right] ^{n-\left\vert i\right\vert }p^{\left\vert i\right\vert }\text{Tr}%
\left\{ \left[ \phi _{A^{\left\{ i\right\} }}\right] ^{\alpha }\right\} %
\bigg] ^{\frac{1}{\alpha }}\right\}  \\
& \leq 2^{-n\left( \frac{\alpha -1}{\alpha }\right) R}\Bigg[ \sum_{i\in
\left\{ 0,1\right\} ^{n}}\left[ \left( 1-p\right) d^{\alpha -1}\right]
^{n-\left\vert i\right\vert }p^{\left\vert i\right\vert }\ \mathbb{E}_{\phi
}\left\{ \text{Tr}\left\{ \bigg[ \phi _{A^{\left\{ i\right\} }}\bigg]
^{\alpha }\right\} \right\} \Bigg]^{\frac{1}{\alpha}},
\label{eq:bound-fid-renyi-1}
\end{align}%
with the first inequality following from the development in the previous
section and the second inequality following from concavity of $x^{\frac{1}{%
\alpha }}$ for $\alpha \in (1,2]$. So it remains to analyze the term
$
\mathbb{E}\left\{ \text{Tr}\left\{ \left[ \phi _{A^{\left\{ i\right\} }}%
\right] ^{\alpha }\right\} \right\} .
$
Let $M_{i}^{\dag }M_{i}=\phi _{A^{\left\{ i\right\} }}$ and consider that%
\begin{align}
\text{Tr}\left\{ [ \phi _{A^{\left\{ i\right\} }}] ^{\alpha
}\right\} & =\text{Tr}\left\{ ( M_{i}^{\dag }M_{i}) ^{\alpha
}\right\}  
 =\text{Tr}\left\{ ( M_{i}^{\dag }M_{i}) ^{\alpha -1}\left(
M_{i}^{\dag }M_{i}\right) \right\}  \\
& \leq \left( \left\Vert M_{i}\right\Vert _{\infty }^{2}\right) ^{\alpha -1}%
\text{Tr}\left\{ \left( M_{i}^{\dag }M_{i}\right) \right\}  
 =\left( \left\Vert M_{i}\right\Vert _{\infty }^{2}\right) ^{\alpha -1}
\end{align}%
By employing the above inequalities and concavity of $x^{\alpha -1}$ for $%
\alpha \in (1,2]$, we find that%
\begin{equation}
\mathbb{E}\left\{ \text{Tr}\left\{ \left[ \phi _{A^{\left\{ i\right\} }}%
\right] ^{\alpha }\right\} \right\} \leq \left[ \mathbb{E}\left\{ \left\Vert
M_{i}\right\Vert _{\infty }^{2}\right\} \right] ^{\alpha -1}.
\end{equation}%
For a randomly chosen pure state $\psi _{RS}$ on systems $R$ and $S$ and
such that $\psi _{R}=M^{\dag }M$, we have the estimate%
\begin{equation}
\mathbb{E}\left\{ \left\Vert M\right\Vert _{\infty }^{2}\right\} \leq
Cd_{R}^{-1},
\end{equation}%
where $d_{R}=\dim \left( \mathcal{H}_{R}\right) $ and $C$ is a universal
constant independent of $d_{R}$ \cite{ASW09}. This then implies the
following bound for our setting:%
\begin{equation}
\mathbb{E}\left\{ \text{Tr}\left\{ \left[ \phi _{A^{\left\{ i\right\} }}%
\right] ^{\alpha }\right\} \right\} \leq \left( Cd^{-\left\vert i\right\vert
}\right) ^{\alpha -1}=C^{\alpha -1}d^{\left\vert i\right\vert \left(
1-\alpha \right) },
\end{equation}%
where we recall that $d$ is the dimension of an individual input to the
channel (so that the support of $\psi _{A^{\left\{ i\right\} }}$ has
dimension $d^{\left\vert i\right\vert }$). Plugging back in to (\ref%
{eq:bound-fid-renyi-1}), we find the upper bound%
\begin{align}
\mathbb{E}_{\phi }\left\{ F\left( \phi \right) \right\} & \leq \left[
2^{-n\left( \frac{\alpha -1}{\alpha }\right) R}\right] \Bigg[ \sum_{i\in
\left\{ 0,1\right\} ^{n}}\left[ \left( 1-p\right) d^{\alpha -1}\right]
^{n-\left\vert i\right\vert }p^{\left\vert i\right\vert }\mathbb{E}\left\{ 
\text{Tr}\left\{ \left[ \phi _{A^{\left\{ i\right\} }}\right] ^{\alpha
}\right\} \right\} \Bigg] ^{\frac{1}{\alpha }} \\
& \leq \left[ 2^{-n\left( \frac{\alpha -1}{\alpha }\right) R}\right] \Bigg[
\sum_{i\in \left\{ 0,1\right\} ^{n}}\left[ \left( 1-p\right) d^{\alpha -1}%
\right] ^{n-\left\vert i\right\vert }p^{\left\vert i\right\vert }C^{\alpha
-1}d^{\left\vert i\right\vert \left( 1-\alpha \right) }\Bigg] ^{\frac{1}{%
\alpha }} \\
& =2^{-n\left( \frac{\alpha -1}{\alpha }\right) R}\ C^{\alpha -1}\ \Bigg[
\sum_{i\in \left\{ 0,1\right\} ^{n}}\left[ \left( 1-p\right) d^{\alpha -1}%
\right] ^{n-\left\vert i\right\vert }\left[ pd^{\left( 1-\alpha \right) }%
\right] ^{\left\vert i\right\vert }\Bigg] ^{\frac{1}{\alpha }} \\
& =2^{-n\left( \frac{\alpha -1}{\alpha }\right) R}\ C^{\alpha -1}\ \left[
\left( 1-p\right) d^{\alpha -1}+d^{1-\alpha }p\right] ^{\frac{n}{\alpha }} \\
& =2^{-n\left( \frac{\alpha -1}{\alpha }\right) \left( R-\frac{1}{\alpha -1}%
\log \left[ \left( 1-p\right) d^{\alpha -1}+d^{1-\alpha }p\right] -\frac{%
\alpha }{n}\log C\right) }.  \label{eq:expected-fidelity-bound}
\end{align}

We now argue that if the rate $R$\ of quantum communication is strictly
larger than the quantum capacity $\left( 1-2p\right) \log d$\ of the erasure
channel, then we can pick $\alpha $ as a constant near one and $n$ large
enough such that%
\begin{equation}
\left( \frac{\alpha -1}{\alpha }\right) \left( R-\frac{1}{\alpha -1}\log %
\left[ \left( 1-p\right) d^{\alpha -1}+d^{1-\alpha }p\right] -\frac{\alpha }{%
n}\log C\right) >0.  \label{eq:positive-exponent}
\end{equation}%
So consider the term:%
\begin{equation}
\frac{1}{\alpha -1}\log \left[ \left( 1-p\right) d^{\alpha -1}+d^{1-\alpha }p%
\right] .
\end{equation}%
Let us set $\alpha =1+t$, so that the above is%
\begin{equation}
\frac{1}{t}\log \left( \left( 1-p\right) d^{t}+d^{-t}p\right) .
\end{equation}%
The limit of this quantity as $t\rightarrow 0$ ($\alpha \rightarrow 1$) is
given by%
\begin{equation}
\Bigg. \frac{\left( 1-p\right) d^{t}\log d-pd^{-t}\log d}{\left( 1-p\right)
d^{t}+d^{-t}p}\Bigg\vert _{t=0}=\left( 1-2p\right) \log d.
\end{equation}%
The other term $-\frac{\alpha }{n}\log C$\ in the exponent becomes
arbitrarily small as $n$ becomes larger. Thus, it is always possible to pick
a constant $\alpha $ and $n$ large enough so that (\ref{eq:positive-exponent}%
) is satisfied, and we recover a strong converse property for the
expectation of the fidelity under randomly chosen entanglement generation
codes.

Since the fidelity $F\left( \phi \right) $\ is a non-negative random
variable between zero and one, we can appeal to Markov's inequality to
recover the following bound:%
\begin{multline}
\Pr_{\phi }\left\{ F\left( \phi \right) >2^{-\frac{1}{2}n\left( \frac{\alpha
-1}{\alpha }\right) \left( R-\frac{1}{\alpha -1}\log \left[ \left(
1-p\right) d^{\alpha -1}+d^{1-\alpha }p\right] -\frac{\alpha }{n}\log
C\right) }\right\}  \\
\leq \frac{\mathbb{E}_{\phi }\left\{ F\left( \phi \right) \right\} }{2^{-%
\frac{1}{2}n\left( \frac{\alpha -1}{\alpha }\right) \left( R-\frac{1}{\alpha
-1}\log \left[ \left( 1-p\right) d^{\alpha -1}+d^{1-\alpha }p\right] -\frac{%
\alpha }{n}\log C\right) }} \\
\leq 2^{-\frac{1}{2}n\left( \frac{\alpha -1}{\alpha }\right) \left( R-\frac{1%
}{\alpha -1}\log \left[ \left( 1-p\right) d^{\alpha -1}+d^{1-\alpha }p\right]
-\frac{\alpha }{n}\log C\right) },
\end{multline}%
where we used the bound in (\ref{eq:expected-fidelity-bound}) for the second
inequality. Thus, our conclusion is that if $R>\left( 1-2p\right) \log d$,
then we can choose $\alpha $ a constant and $n$ large enough so that (\ref%
{eq:positive-exponent}) holds, with the fraction of codes satisfying the
strong converse property rapidly approaching one as the number of channel
uses increases.

We can obtain an even sharper statement about the convergence by appealing
to Levy's Lemma (see \cite{F12}, for example):

\begin{lemma}
[Levy's Lemma]Let $f:\mathbb{C}^{d}\rightarrow\mathbb{R}$ and $\eta>0$ be such
that for all pure states $\left\vert \varphi_{1}\right\rangle $ and
$\left\vert \varphi_{2}\right\rangle $ in $\mathbb{C}^{d}$%
\[
\left\vert f\left(  \left\vert \varphi_{1}\right\rangle \right)  -f\left(
\left\vert \varphi_{1}\right\rangle \right)  \right\vert \leq\eta\left\Vert
\left\vert \varphi_{1}\right\rangle -\left\vert \varphi_{2}\right\rangle
\right\Vert _{2}.
\]
Let $\left\vert \varphi\right\rangle $ be a random pure state in
$\mathbb{C}^{d}$. Then for all $\delta\in\left[  0,\eta\right]  $, the
following bound holds%
\[
\Pr\left\{  \left\vert f\left(  \left\vert \varphi\right\rangle \right)
-\mathbb{E}\left\{  f\left(  \left\vert \varphi\right\rangle \right)
\right\}  \right\vert \geq\delta\right\}  \leq4\exp\left\{  -\frac{d\delta
^{2}}{c\eta}\right\}  ,
\]
where $c$ is a positive constant.
\end{lemma}

We obtain a Lipschitz constant for the fidelity as a function of pure input
states as follows:%
\begin{align}
\left\vert F\left( \varphi _{1}\right) -F\left( \varphi _{2}\right)
\right\vert & \leq \left\vert F\left( \varphi _{1}\right) -F\left( \varphi
_{2}\right) \right\vert +\left\vert \left[ 1-F\left( \varphi _{1}\right) %
\right] -\left[ 1-F\left( \varphi _{2}\right) \right] \right\vert  \\
& \leq \left\Vert \varphi _{1}-\varphi _{2}\right\Vert _{1} \\
& \leq 2\left\Vert \left\vert \varphi _{1}\right\rangle -\left\vert \varphi
_{2}\right\rangle \right\Vert _{2}.
\end{align}%
The first inequality is obvious, the second follows from monotonicity of
trace distance under quantum operations (with these operations being a test
for the maximally entangled state, the decoder, the channel and the
encoder), and the third inequality is straightforward (see Lemma~I.4 in \cite%
{D09}, for example).

Since we have the bound%
\begin{equation}
0\leq \mathbb{E}_{\phi }\left\{ F\left( \phi \right) \right\} \leq
2^{-n\left( \frac{\alpha -1}{\alpha }\right) \left( R-\frac{1}{\alpha -1}%
\log \left[ \left( 1-p\right) d^{\alpha -1}+d^{1-\alpha }p\right] -\frac{%
\alpha }{n}\log C\right) }\equiv g,
\end{equation}%
it follows from Levy's lemma that%
\begin{align}
\Pr \left\{ F\left( \phi \right) \geq g+\delta \right\} & \leq \Pr \left\{
F\left( \phi \right) \geq \mathbb{E}_{\phi }\left\{ F\left( \phi \right)
\right\} +\delta \right\}  \\
& \leq 4\exp \left\{ -\frac{2^{n\left[ R+\log d\right] }\delta ^{2}}{2c}%
\right\} 
\end{align}%
We can take $\delta =g$, to find that%
\begin{multline}
\Pr \left\{ F\left( \phi \right) \geq 2\cdot 2^{-n\left( \frac{\alpha -1}{%
\alpha }\right) \left( R-\frac{1}{\alpha -1}\log \left[ \left( 1-p\right)
d^{\alpha -1}+d^{1-\alpha }p\right] -\frac{\alpha }{n}\log C\right)
}\right\}  \\
\leq 4\exp \left\{ -\frac{2^{n\left[ R+\log d\right] }\big[ 2^{-n\left( 
\frac{\alpha -1}{\alpha }\right) \left( R-\frac{1}{\alpha -1}\log \left[
\left( 1-p\right) d^{\alpha -1}+d^{1-\alpha }p\right] -\frac{\alpha }{n}\log
C\right) }\big] ^{2}}{2c}\right\} .
\end{multline}%
Now, without loss of generality, we can take $R\leq \log d$ (otherwise the
strong converse already holds for all codes), so that $R+\log d\geq 2\left(\frac{%
\alpha -1}{\alpha }\right)R$. Thus, we see that the fraction of codes with $%
R>\left( 1-2p\right) \log d$ and obeying the strong converse approaches one
doubly exponentially fast in the number of channel uses.

\section{Conclusion}

The main result of the present paper is a proof that the large fraction of
codes with a quantum communication rate exceeding the quantum capacity of
the erasure channel satisfy the strong converse. We view this result as
adding to the evidence from \cite{MW13}\ that a strong converse should hold
for the quantum capacity of these channels. The main open question going
forward from here is to prove that a fully strong converse holds for the
quantum capacity of the erasure channel (i.e., that if the rate of any
quantum communication scheme exceeds the quantum capacity of the erasure
channel, then the quantum error necessarily converges to one).

%PUT HERE REMARK ON $p=\frac12$ ERASURE CHANNEL? 
%($\sqrt{n}$ TERM CONSISTENT WITH LOW FIDELITY...)

\bigskip
The focus on the erasure channel of the present discussion may be justified
by the simplicity of the channel (including its additivity). It also allowed
us to give an
illustration of the power of the R\'{e}nyi divergence approach.
At the same time, it seems to be true for all currently known random
code ensembles achieving the coherent information for a channel
$\mathcal{N}$ with Stinespring isometry $V:A' \hookrightarrow B \otimes E$ 
(with respect to a given input density $\rho_A$), 
that at rates above the same coherent
information they have fidelity going to zero, with overwhelming probability.
Of course this has to be verified for each ensemble separately, but
rests on two properties that hold for most codes in the ensemble.
Namely, with respect to the pure state 
$\ket{\psi}_{A B^nE^n} = (I \otimes V^{\otimes n})\ket{\phi}_{A{A'}^n}$:
\begin{enumerate}
  \item {\bf Typicality of $\mathbf{B}$}. The channel output $\psi_{B^n}$
    is largely in the typical subspace of $\cN(\rho_A)^{\otimes n}$ in the
    sense that $H_{\max}^\delta(B^n) \leq n S(\cN(\rho_A)) + o(n)$.

  \item {\bf Saturation of $\mathbf{E}$}. The complementary channel output
    $\psi_{E^n}$ covers essentially uniformly the typical subspace of
    ${\cN^c}(\rho_A)^{\otimes n}$ in the sense that 
    $H_{\min}^\delta(E^n) \geq n S(\cN^c(\rho_A)) - o(n)$.
\end{enumerate}
[In fact, in practice the latter property tends to be true for 
most states in most code subspaces.]
We refer to \cite{T12} (cf.~\cite{MW13}) for the definitions and necessary
properties of (smooth) min- and max-entropies used in the following. 

Now, if our code is supposed to generate
entanglement at rate $R$ with fidelity $F$, then by the decoupling
principle,
\begin{equation}
  \label{eq:min-lower}
  H_{\min}^{\sqrt{1-F^2}}(A|E^n) \geq nR.
\end{equation}
On the other hand, using relations between min- and max-entropies
as well as chain rules,
\begin{equation}\begin{split}
  \label{eq:min-upper}
  H_{\min}^{\sqrt{1-F^2}}(A|E^n) &\lesssim H_{\max}^{\epsilon}(A|E^n) \\
                                 &\lesssim H_{\max}^{\delta}(AE^n) - H_{\min}^{\delta}(E^n) \\
                                 &=        H_{\max}^{\delta}(B^n) - H_{\min}^{\delta}(E^n),
\end{split}\end{equation}
where $\epsilon = \frac12(1-\sqrt{1-F^2})$ and $\delta = \frac14\epsilon$,
the inequalities are true up to terms of order $\log\frac{1}{\delta}$.
By the typicality and saturation properties, (\ref{eq:min-lower}) and
(\ref{eq:min-upper}) bound the rate as desired,
\begin{equation}
  R \leq S(\cN(\rho_A)) - S(\cN^c(\rho_A)) + o(1)
    =    I(A\rangle B) + o(1).
\end{equation}

\bigskip
\textbf{Acknowledgements.} We are grateful to Naresh Sharma for many
conversations from which the ideas in this paper arose. We thank the Isaac
Newton Institute for Mathematical Sciences at the University of Cambridge
for organizing the semester ``Mathematical Challenges in Quantum
Information,'' at which we had an opportunity to discuss this research. MMW
is grateful to the Department of Physics and Astronomy at Louisiana State
University for startup funds that supported this research and acknowledges
support from the DARPA\ Quiness Program through US\ Army Research Office
award W31P4Q-12-1-0019. AW acknowledges financial support by the Spanish
MINECO, project FIS2008-01236 with the support of FEDER funds, the EC STREP
\textquotedblleft RAQUEL\textquotedblright, the ERC Advanced Grant
\textquotedblleft IRQUAT\textquotedblright, and the Philip Leverhulme Trust.

\appendix

\section{Strong converse for the classical capacity of the quantum erasure
channel}

In this appendix, we detail a proof that the strong converse holds for the
classical capacity of the quantum erasure channel. To our knowledge, a proof
of this statement has not yet appeared in the literature. This result was obtained
in collaboration with Naresh Sharma.

Using the generalized divergence framework established in \cite{SW12} and
reviewed in \cite{WWY13}\ (or even the method of Koenig-Wehner \cite{KW09}),
we obtain the following bound on the success probability when transmitting a
classical message through the quantum erasure channel%
\begin{equation}
p_{\text{succ}}\leq 2^{-n\left( \frac{\alpha -1}{\alpha }\right) \left( R-%
\frac{1}{n}\chi _{\alpha }\left( \mathcal{N}^{\otimes n}\right) \right) },
\label{eq:success-prob-bound}
\end{equation}%
where%
\begin{equation}
\frac{1}{n}\chi _{\alpha }\left( \mathcal{N}^{\otimes n}\right) 
\end{equation}%
is the regularized R\'{e}nyi-Holevo information of the erasure channel. So
our goal is to prove that this quantity is additive as a function of the
quantum erasure channel. First recall that this quantity can be written as
an information radius \cite{SW01,WWY13}:%
\begin{equation}
\chi _{\alpha }\left( \mathcal{N}^{\otimes n}\right) =\min_{\sigma
_{B^{n}}}\max_{\rho _{A^{n}}}D_{\alpha }\left( \mathcal{N}^{\otimes n}\left(
\rho _{A^{n}}\right) ||\sigma _{B^{n}}\right) .
\end{equation}%
With this, we see that we can upper bound this quantity simply by choosing $%
\sigma _{B^{n}}$ to be the output of the erasure channel when the
tensor-power maximally mixed state is input:%
\begin{equation}
\chi _{\alpha }\left( \mathcal{N}^{\otimes n}\right) \leq \max_{\rho
_{A^{n}}}D_{\alpha }\left( \mathcal{N}^{\otimes n}\left( \rho
_{A^{n}}\right) ||\left[ \mathcal{N}\left( \pi \right) \right] ^{\otimes
n}\right) .
\end{equation}%
As discussed in Section~\ref{sec:erasure-channel-app}, the output of the
quantum erasure channel is rather special, in the sense that it can be
written as a linear combination of $2^{n}$ density operators which are supported on
orthogonal subspaces. We can index these by a binary string $i$ (where ones
in this string represent the systems that get erased and zeros represent
systems that do not get erased), and we denote the density operators for $%
\mathcal{N}^{\otimes n}\left( \rho _{A^{n}}\right) $ by $\omega _{B^{n}}^{i}$
and those for $\left[ \mathcal{N}\left( \pi \right) \right] ^{\otimes n}$ by 
$\tau _{B^{n}}^{i}$. Furthermore, let $\left\{ i\right\} \ $be the set of
indices for the systems that get erased, so that we denote the systems that
get erased by $A^{\left\{ i\right\} }$ and those that do not by $A^{\left\{
i\right\} ^{c}}$. We then find that%
\begin{align}
& \max_{\rho _{A^{n}}}D_{\alpha }\left( \mathcal{N}^{\otimes n}\left( \rho
_{A^{n}}\right) ||\left[ \mathcal{N}\left( \pi \right) \right] ^{\otimes
n}\right)   \nonumber \\
& =\frac{1}{\alpha -1}\log \max_{\rho _{A^{n}}}\text{Tr}\left\{ \left[ 
\mathcal{N}^{\otimes n}\left( \rho _{A^{n}}\right) \right] ^{\alpha }\left( %
\left[ \mathcal{N}\left( \pi \right) \right] ^{\otimes n}\right) ^{1-\alpha
}\right\}  \\
& =\frac{1}{\alpha -1}\log \max_{\rho _{A^{n}}}\sum_{i\in \left\{
0,1\right\} ^{n}}\left( 1-p\right) ^{n-\left\vert i\right\vert
}p^{\left\vert i\right\vert }\text{Tr}\left\{ \left[ \omega _{B^{n}}^{i}%
\right] ^{\alpha }\left[ \tau _{B^{n}}^{i}\right] ^{1-\alpha }\right\}  \\
& =\frac{1}{\alpha -1}\log \max_{\rho _{A^{n}}}\sum_{i\in \left\{
0,1\right\} ^{n}}\left( 1-p\right) ^{n-\left\vert i\right\vert
}p^{\left\vert i\right\vert }\text{Tr}\left\{ \left[ \rho _{A^{\left\{
i\right\} ^{c}}}\right] ^{\alpha }\left[ \pi _{A^{\left\{ i\right\} ^{c}}}%
\right] ^{1-\alpha }\right\} 
\end{align}%
The above equalities follow simply by substitution and some algebra.
Continuing, the last line above is equal to%
\begin{align}
& =\frac{1}{\alpha -1}\log \max_{\rho _{A^{n}}}\sum_{i\in \left\{
0,1\right\} ^{n}}\left[ \left( 1-p\right) d^{\alpha -1}\right]
^{n-\left\vert i\right\vert }p^{\left\vert i\right\vert }\text{Tr}\left\{ %
\left[ \rho _{A^{\left\{ i\right\} ^{c}}}\right] ^{\alpha }\right\}  \\
& \leq \frac{1}{\alpha -1}\log \sum_{i\in \left\{ 0,1\right\} ^{n}}\left[
\left( 1-p\right) d^{\alpha -1}\right] ^{n-\left\vert i\right\vert
}p^{\left\vert i\right\vert } \\
& =\frac{1}{\alpha -1}\log \sum_{k=0}^{n}\left[ \left( 1-p\right) d^{\alpha
-1}\right] ^{n-k}p^{k}\binom{n}{k} \\
& =\frac{1}{\alpha -1}\log \left( \left( 1-p\right) d^{\left( \alpha
-1\right) }+p\right) ^{n} \\
& =n\left[ \frac{1}{\alpha -1}\log \left( \left( 1-p\right) d^{\left( \alpha
-1\right) }+p\right) \right] 
\end{align}%
The inequality follows because Tr$\left\{ \left[ \rho _{A^{\left\{ i\right\}
^{c}}}\right] ^{\alpha }\right\} \leq 1$ for all $\alpha \geq 1$ (and we are
considering $\alpha \in (1,2]$ here). The next few equalities are
straightforward. Returning to (\ref{eq:success-prob-bound}), all of this
development implies that we get the following upper bound on success
probability%
\begin{equation}
p_{\text{succ}}\leq 2^{-n\left( \frac{\alpha -1}{\alpha }\right) \left( R-%
\left[ \frac{1}{\alpha -1}\log \left( \left( 1-p\right) d^{\left( \alpha
-1\right) }+p\right) \right] \right) }
\end{equation}

The last line above is a single-letter upper bound. Now, let us set $\alpha
=1+t$, so that the above is%
\begin{equation}
\frac{1}{t}\log \left( \left( 1-p\right) d^{t}+p\right) .
\end{equation}%
The limit of this quantity as $t\rightarrow 0$ is given by%
\begin{equation}
\left. \frac{\left( 1-p\right) d^{t}\log d}{\left( 1-p\right) d^{t}+p}%
\right\vert _{\varepsilon =0}=\left( 1-p\right) \log d,
\end{equation}%
which is exactly the classical capacity of the quantum erasure channel.
Thus, whenever the classical communication rate $R>\left( 1-p\right) \log d$%
, we can always find a value of $\alpha $ in a neighborhood of one such that%
\begin{equation}
\left( \frac{\alpha -1}{\alpha }\right) \left( R-\left[ \frac{1}{\alpha -1}%
\log \left( \left( 1-p\right) d^{\left( \alpha -1\right) }+p\right) \right]
\right) >0.
\end{equation}%
This concludes the proof.

Interestingly, the proof above demonstrates that tensor-product pure-state
codewords are the optimal choice in order to saturate the bound given above.
That is, for pure-state codewords, we have the equality Tr$\left\{ \left[
\rho_{A^{\left\{ i\right\} ^{c}}}\right] ^{\alpha}\right\} =1$, so that the
upper bound is saturated by this choice.

\bibliography{Ref}

\end{document}